\documentclass[12pt]{article}
\usepackage{graphicx}
\usepackage{amsmath}
\usepackage{amssymb}
\usepackage{array}
\usepackage{cite}
\usepackage{epsfig}
\usepackage[english]{babel}
\usepackage[latin1]{inputenc}
\usepackage[T1]{fontenc}
\usepackage{amsfonts}
\usepackage{float}
\usepackage{sidecap}
\usepackage{amsbsy}
\usepackage[margin=10pt,labelfont=bf]{caption}
\usepackage{multirow}
\usepackage{longtable}
\usepackage{color}
\usepackage{lscape}
\usepackage[dvips,pdftitle={The Web of D-branes at Singularities in Compact Calabi-Yau Manifolds}, pdfauthor={Michele Cicoli, Sven Krippendorf, Christoph Mayrhofer, Fernando Quevedo, Roberto Valandro}, pdfsubject={string phenomenology}, pdfkeywords={}]{hyperref}
\usepackage{collref}

\newcommand{\bea}{\begin{eqnarray}}
\newcommand{\eea}{\end{eqnarray}}

\newcommand{\be}{\begin{equation}}
\newcommand{\ee}{\end{equation}}
\newcommand{\bi}{\begin{itemize}}
\newcommand{\ei}{\end{itemize}}
\newcommand{\ben}{\begin{enumerate}}
\newcommand{\een}{\end{enumerate}}

\def\ba{\begin{eqnarray}}
\def\ea{\end{eqnarray}}
\def\nn{\nonumber}

\makeatletter
\def\x@arrow{\DOTSB\Relbar}
\def\xlongequalsignfill@{\arrowfill@\x@arrow\Relbar\x@arrow}
\newcommand{\xlongequal}[2]{%
    \ext@arrow 0099\xlongequalsignfill@{#1}{#2}}
\makeatother

\newcommand{\roughly}[1]{\mathrel{\raise.3ex\hbox{$#1$\kern-0.85em
\lower1ex\hbox{$\sim$}}}}

\def\endignore{}
\def\ignore #1\endignore{}
\def\nn{\nonumber}

\def\beq{\begin{equation}}
\def\eeq{\end{equation}}
\def\beqa{\begin{eqnarray}}
\def\eeqa{\end{eqnarray}}

\def\cC{{\cal C}}
\def\cD{{\cal D}}
\def\cE{{\cal E}}
\def\cF{{\cal F}}

\def\cN{{\cal N}}

\def\cV{{\cal V}}

\newcommand{\bmat}{\left(\begin{array}}
\newcommand{\emat}{\end{array}\right)}

\def\endignore{}
\def\ignore #1\endignore{}

\def\-{\hphantom{-}}

\def\s2{\frac{1}{2}}

\def\diag{{\rm diag \,}}

\def\IF{\relax{\rm I\kern-.18em F}}
\def\II{\relax{\rm I\kern-.18em I}}
\def\IP{\relax{\rm I\kern-.18em P}}
\def\IC{\relax{\rm I\kern-.48em C}}
\def\IR{\relax{\rm I\kern-.18em R}}
\def\IK{\relax{\rm I\kern-.20em K}}
\def\IM{\relax{\rm I\kern-.25em M}}

\def\Dsl{\,\raise.15ex\hbox{/}\mkern-13.5mu D}

\def \one{\relax{\rm 1\kern-.26em I}}

\def\nn{\nonumber}

\def\({\left(}
\def\){\right)}

\begin{document}
\begin{flushright} DAMTP-2013-17\\ \hspace{1.5cm}
\end{flushright}
\begin{center}
{\bf\LARGE The Web of D-branes at Singularities\\ \vspace{3mm} in Compact Calabi-Yau Manifolds}
\\[0.81cm]
Michele Cicoli${}^{1,2,3},$ Sven Krippendorf${}^4,$ Christoph Mayrhofer${}^5,$ \\ Fernando Quevedo${}^{3,6},$ Roberto Valandro${}^{3,7}$\\[0.3cm]

{\it \small $^1$ Dipartimento di Fisica e Astronomia, Universit\`a di Bologna, \\ via Irnerio 46, 40126 Bologna, Italy. \\
$^2$ INFN, Sezione di Bologna, Italy.\\
$^3$ ICTP, Strada Costiera 11, Trieste 34014, Italy.\\
$^4$ Bethe Center for Theoretical Physics and Physikalisches Institut der\\ Universit\"at Bonn, Nussallee 12, 53115 Bonn, Germany.\\
$^5$ Institut f\"ur Theoretische Physik, Universit\"at Heidelberg,\\ Philosophenweg 19,  69120 Heidelberg, Germany.\\
$^6$ DAMTP, University of Cambridge, Wilberforce Road, Cambridge, CB3 0WA, UK.\\
$^7$ INFN, Sezione di Trieste, Italy.}\\[0.5cm]

\date{\today}
\end{center}
\begin{abstract}
\noindent

We present novel continuous supersymmetric transitions which take place among different chiral configurations of D3/D7 branes at singularities in the context of type IIB Calabi-Yau compactifications. We find that distinct local models which admit a consistent global embedding can actually be connected to each other along flat directions by means of transitions of bulk-to-flavour branes. This has interesting interpretations in terms of brane recombination/splitting and brane/anti-brane creation/annihilation. These transitions give rise to a large web of quiver gauge theories parametrised by splitting/recombination modes of bulk branes which are not present in the non-compact case. We illustrate our results in concrete global embeddings of chiral models at a dP$_0$ singularity.

\end{abstract}

\thispagestyle{empty}
\clearpage
\setcounter{page}{1}

\newpage
\tableofcontents

\section{Introduction and summary}
\label{Introduction}

The moduli spaces of supersymmetric string constructions have proven to be far richer than expected. In fact, there are several examples of string models which were initially thought to be unrelated to each other but after a detailed investigation of their moduli space turned out to be continuously connected. A primary example is given by  Calabi-Yau  (CY) compactifications  which are believed to be all connected to each other by topology changing conifold transitions (for a review see for instance \cite{Greene:1996cy}).

Further examples have been discovered independently over the years. In~\cite{Ibanez:1987xa,Font:1988tp}, working in the context of heterotic orbifold compactifications, different chiral models obtained by discrete Wilson lines were found to be actually connected by continuous Wilson lines which lower the rank of the gauge group.
The stringy nature of this phenomenon is manifested by the impossibility to describe it in the effective field theory
language in terms of a flat direction in a single patch of field space.

Afterwards, Kachru and Silverstein showed that four-dimensional $\cN=1$ heterotic vacua with different chiral content can be connected by
phase transitions which can be described in the Horava-Witten M-theory picture as M5-branes moving away
from one $E_8$ plane \cite{hep-th/9704185}.
Subsequently, Douglas and Zhou pointed out that  
vacua with different gauge group and chiral spectrum are also connected
at the classical level via continuous transitions among different supersymmetric configurations \cite{hep-th/0403018}.
They illustrated their general claims in the case of heterotic compactifications on smooth CY three-folds
where these chirality changing transitions take place via deformations of the gauge bundle (parameterised by charged
bundle moduli) and in the case of type IIA CY orientifolds with D6-branes at angles where continuous deformations
of the three-cycles wrapped by the branes (parameterised by open string moduli corresponding to splitting/recombination modes)
drive transitions between vacua with different gauge groups and chiral matter.
Notice that these transitions occur without going through a potential barrier since they
connect two supersymmetric configurations with the same central charge.

In this paper, we extend these results uncovering a novel transitions in which  chirality changing transitions can
take place
in the case of type IIB compactifications on CY orientifolds,
focusing on models with fractional branes at singularities. Following our previous works on
chiral local model building in explicit compact CY backgrounds \cite{Cicoli:2011qg,Cicoli:2012vw},
in a recent paper \cite{Cicoli:2013mpa} we provided a consistent global embedding of generic
local models involving both fractional D3- and `flavour' D7-branes at singularities as well as bulk D7-branes
wrapping divisors which do not intersect the singularity.
The analysis of \cite{Cicoli:2013mpa} revealed that not all models
which can be built locally, admit a consistent global embedding.
In this paper, we shall complete this analysis, showing that those models
which can be realised globally are actually  continuously connected
to each other by supersymmetric transitions involving D7-branes coming from the bulk.
In this regard, models which appeared to be completely disconnected from the local point of view,
turn out to be all related from the global point of view giving rise to a `web of quiver gauge theories'
parameterised by open string moduli corresponding to splitting/recombination modes of bulk D7-branes.

Let us summarise the main features of these supersymmetric transitions focusing on the case of branes
at dP$_0=\mathbb{P}^2$ singularities, i.e.~at $\mathbb{C}^3/\mathbb{Z}_3$ orbifold singularities. We start by recalling that, in this case,
there are three different kinds of fractional D3-branes, that we call $D3_{\rm frac}^0$, $D3_{\rm frac}^1$ and $D3_{\rm frac}^2$, with different D7-, D5- and D3-charges~\cite{Diaconescu:1999dt,Douglas:2000qw}.
In the resolved picture, when the dP$_0$ divisor is blown up, the fractional branes can be seen as D7/$\overline{{\rm D7}}$ branes wrapping the blown-up dP$_0$ and supporting different gauge bundles.\footnote{When the singularity is resolved, the fractional branes are not stable anymore, as supersymmetry is broken. Anyway, the resolved picture is usefull to compute quantities that do not depend on the smoothing parameter (in this case the K\"ahler modulus controlling the resolution).}
In particular, the fractional branes $D3_{\rm frac}^{0,2}$ appear as fluxed $\overline{{\rm D7}}$-branes,\footnote{The fractional branes
$D3_{\rm frac}^{0}$ and $D3_{\rm frac}^{2}$ are distinguished by the different gauge bundles living on them, as we will explain in Section~\ref{sec:BPS_D-branes} (cf.~\eqref{ChFi}).} 
 while the fractional brane $D3_{\rm frac}^1$ as a rank-two D7-brane stack with a non-Abelian gauge bundle. 

The fact that the fractional D3-branes can carry either negative ($D3_{\rm frac}^{0,2}$) or positive ($D3_{\rm frac}^1$) D7-charge allows
two different kinds of basic bulk-to-flavour D7-brane transitions.
\bi
\item \textbf{Transitions of type I}

In the singular space, these transitions take place via the following `reaction':
\be
D7_{\rm bulk}^{0,2} + D3_{\rm frac}^{0,2} \quad \leftrightarrow \quad D7_{\rm flav}^{0,2}
\label{TransI}
\ee
which occurs when a bulk D7-brane not passing through the singularity is deformed continuously until it
touches the singularity. At this point, this bulk D7-brane combines with a fractional brane of type $0$ or $2$
to give rise to a flavour D7-brane.
We have denoted bulk and flavour D7-branes as of type $0$ or $2$ depending on which fractional brane is involved in the transition.
This process reduces the rank of the gauge
group at a node of the quiver by one  and adds one flavour brane. 
Moreover, even if the number of massless chiral modes does not change, these fields
get rearranged in different representations of the new gauge theory.
Notice also that the arrows in (\ref{TransI}) go in both directions, meaning that this process can also increase the rank
of the gauge group and decrease the number of flavour branes.

In the resolved space (when the dP$_0$ divisor has finite size), this transition can be understood by the following steps:
\ben
\item A bulk D7-brane splits into a flavour D7-brane and a D7-brane wrapping the dP$_0$ divisor.

\item The D7-brane on dP$_0$ annihilates with one fractional brane carrying the charge of an $\overline{\rm D7}$-brane, i.e.\ $D3_{\rm frac}^{0}$ or $D3_{\rm frac}^{2}$ depending on the gauge bundle on the brane wrapping the dP$_0$.
\een
This process can be summarised as follows
\be
D7_{\rm bulk}^{0,2} + D3_{\rm frac}^{0,2} 
\underset{\rm splitting/recombination}{\longleftrightarrow} D7_{\rm flav}^{0,2} + D7_{{\rm on \,dP}_0}^{0,2} + D3_{\rm frac}^{0,2} 
\underset{\rm annihilation/creation}{\longleftrightarrow} D7_{\rm flav}^{0,2} \, .\nn
\ee
Notice that again the transition can take place in both directions through brane splitting/recombination and brane annihilation/creation processes.

\item \textbf{Transitions of type II}

In the singular space, the following `reaction' describes these transitions
\be
2\, D7_{\rm bulk}^1  \quad \leftrightarrow \quad D7_{\rm flav}^0 + D7_{\rm flav}^2 + D3_{\rm frac}^1
\label{TransII}
\ee
where a rank-two bulk D7-brane is deformed continuously until it touches the singularity
and then transforms into two flavour D7-branes and a fractional D3-brane of type $1$.\footnote{
In principle, one can also consider a parent transition, in which one bulk brane $D7^1_{\rm bulk}$ splits into a fractional brane $D3_{\rm frac}^1$ and one flavour brane $D7_{\rm flav}^1$. This is related to the described transition of type II by D7-brane recombination.}
This process increases both the rank of the gauge
group at one node of the quiver and the number of flavour branes.
Furthermore, contrary to transitions of type I, 
the number of massless chiral fields gets modified. 
The arrows in \eqref{TransII} go again in both directions, since this transition can also decrease both the rank
of the gauge group and the number of flavour branes.

From the resolved point of view, this transition can be understood as due to the splitting of a
rank-two bulk D7-brane into two flavour D7-branes and a rank-two D7-brane wrapping the dP$_0$ divisor. In order for this stack to play the r\^ole of a type $1$ fractional D3-brane, it needs to support a non-Abelian bundle.
As we will see in the explicit example in Section~\ref{sec:example}, in order for this transition to occur as presented in \eqref{TransII}, the initial bulk D7-brane, denoted as of type $1$,
has to carry a non-Abelian gauge flux as well.
This process can be summarised as follows
\be
2\,D7_{\rm bulk}^1 \underset{\rm splitting/recombination}{\longleftrightarrow}
D7_{\rm flav}^0 + D7_{\rm flav}^2 + 2\,D7_{{\rm on \,dP}_0}^1
\equiv D7_{\rm flav}^0 + D7_{\rm flav}^2 + D3_{\rm frac}^1\, . \nn
\ee
Notice that again the transition can take place in both directions through brane splitting/recombination processes.
\ei

Following this description of supersymmetric transitions, one finds a different way to realise 
the process of removing a D3-brane from a singularity,
i.e.~recombining three fractional D3-branes into a mobile D3-brane in the bulk
\be
D3_{\rm frac}^0 + D3_{\rm frac}^1+ D3_{\rm frac}^2\quad \leftrightarrow \quad D3_{\rm mobile}\, .
\label{RemoveD3}
\ee
This process can be effectively understood as the result of a combination of two transitions of type I
with one transition of type II which take place simultaneously. In fact, combining two reactions of type (\ref{TransI})
and one reaction of type (\ref{TransII}) one obtains:
\be
D7_{\rm bulk}^0 + D7_{\rm bulk}^2 + D3_{\rm frac}^0 + D3_{\rm frac}^1 + D3_{\rm frac}^2 \,\, \underset{\rm type\, I}{\longleftrightarrow} \,\,
D7_{\rm flav}^0 + D7_{\rm flav}^2 + D3_{\rm frac}^1 \,\, \underset{\rm type\, II}{\longleftrightarrow} \,\,
2\, D7_{\rm bulk}^1 \nn
\ee
which reproduces exactly the transition (\ref{RemoveD3}) by considering the bulk branes as spectators.
In a further step a mobile D3-brane is generated by a flux/brane transition, through which a D3-brane
living on the rank-two stack world-volume is expelled into the CY background \cite{Douglas:1995bn}:
\be
2\, D7_{\rm bulk}^1 \quad \underset{\rm flux/brane}{\longleftrightarrow} \quad D7_{\rm bulk}^0 + D7_{\rm bulk}^2 + D3_{\rm mobile}\, .
\ee

Notice that these processes are energy conserving since they correspond to transitions among different BPS configurations with the same charge. In fact, this `web of quiver gauge theories' is parameterised by flat directions corresponding to the splitting and recombination modes of bulk D7-branes. These are open string modes living on bulk D7-branes which are not taken into account in the non-compact case. This explains why these different quiver gauge theories which seem completely unrelated from the local point of view are connected in the global embedding.\footnote{In the local perspective, some transitions not involving bulk D7 branes seem visible, see for example~\cite{Feng:2002fv,Hanany:2001py}.}
In this paper we describe the \emph{kinematics} of transitions among different quiver gauge theories
but in order to understand also the \emph{dynamics} of these transitions,
one would need to explore how to lift these open string flat directions in order to select a vacuum in this landscape of quiver gauge theories.

We stress that these different gauge theories are connected by transitions which are really continuous transformations
contrary to transitions among different string vacua which involve discrete variations of underlying parameters like
the angles between branes in type IIA (which take discrete values because of the periodicity of the tori)
or the gauge fluxes in type IIB (which are integers because of their quantisation condition).

This paper is organised as follows. In Section~\ref{sec:BPS_D-branes} we briefly review the stability conditions for BPS D-branes
at the singular locus and in the geometric regime, considering quiver gauge theories at dP$_0$ singularities
both in the non-compact and in the compact case. In Section~\ref{sec:transition} we provide a general description of the kinematics
of transitions among different supersymmetric brane configurations at dP$_0$ singularities whereas
Section~\ref{sec:example} is devoted to the illustration of these general claims for the explicit construction
carried out in \cite{Cicoli:2013mpa}. We will end this paper with our conclusions in Section~\ref{sec:conclusions}.

\section{D-branes at dP$_0$ singularities}
\label{sec:BPS_D-branes}

\subsection{BPS D-branes}

We consider type IIB string theory compactified on a CY three-fold.
A D$p$-brane is a $(p+1)$-dimensional object which, when space-time filling, wraps a $(p-3)$-cycle $D$ of the compact CY $X$ and is characterised by the choice of the vector bundle $\cE$ over $D$.
A Dp-brane is charged under the RR $(p+1)$-potential. The RR charges of a D-brane are encoded in the `Mukai' charge vector $\Gamma_{\cE,D}$
\begin{equation}
 \Gamma_{{\cal E},D} = D\wedge \mbox{ch} ({\cal E}) \wedge \sqrt{\frac{\mbox{Td}(TD)}{\mbox{Td}(ND)}} \:.
\label{ChargeVectD7br}
\end{equation}
$D$ is the Poincar\'e dual form to the $(p-3)$-cycle in $X$. Td$(V)= 1 + \frac12 c_1(V) + \frac{1}{12}(c_1(V)^2+c_2(V))+... $ is the Todd class of the vector bundle $V$,
$TD$ is the tangent bundle of $D$ and $ND$ the normal bundle of $D$ in $X$ while
ch$({\cal E})$ is the Chern character of the vector bundle ${\cal E}$ (or sheaf) living on the brane.\footnote{The charge vector can also be written in terms of the A-roof genus $\hat{A}$, by shifting the sheaf $\cE$ to the sheaf ${\cal W}={\cal E}\otimes K_S^{1/2}$ whose first Chern class is identified with the gauge flux.}
The D9-charge is encoded in the  zero-form component of $e^{-B}\Gamma_{\cal E}$,  the D7-charge in the two-form,
the D5-charge in the four-form and the D3-charge in the six-form.\footnote{These p-forms are actually the `push-forwards' to the CY manifold $X$ of forms on the D-brane (for a review see \cite{Aspinwall:2004jr}). For that reason, a two-form flux on a D7-brane, Poincar\'e dual to a curve $C$ whose push-forward is trivial on $X$ but non-trivial on the D7-brane, will appear in the D3-charge (six-form) but not in the D5-charge (four-form).}

At large radius of the cycle $D$, the BPS condition on the D-brane is that the wrapped cycle $D$ is holomorphic and the vector bundle satisfies the Hermitian Yang-Mills equation, or equivalently that it
is a holomorphic and stable bundle. These are the F-flatness and D-flatness conditions.
The correct and most general mathematical definition for a (B-type) BPS D-brane in type IIB string theory is that it is a $\Pi$-stable object in the derived category of coherent sheaves on $X$ \cite{Douglas:2000gi}.
The $\Pi$-stability condition \cite{Douglas:2000ah} is a condition on the central charge $Z(\cE)$ of the D-brane: define the `grade' of a D-brane as $$\varphi(\cE)=\frac{1}{\pi} \arg Z(\cE) =\frac{1}{\pi} \textmd{Im}\,\log Z(\cE)\,.$$
The D-brane $\cE$ is $\Pi$-stable if for all sub-objects  $\cE'\subset \cE$ one has $\varphi(\cE')\leq \varphi(\cE)$.

The central charge $Z(\cE)$ of a (B-type) BPS D-brane depends purely on the complexified K\"ahler form $B+i\,J$ and it is independent of the complex structure moduli of $X$. When the D-brane wraps a large cycle the central charge is, up to quantum corrections,
\begin{equation}\label{eq:central-charge_D-brane}
Z(\cE)=\int_X e^{-B-iJ}\Gamma_{\cE,D}\,.
\end{equation}
From \eqref{eq:central-charge_D-brane} we see that for a D3-brane at a generic point of $X$ we have $Z=-1$.

The central charge and, consequently, the stability condition depend on the K\"ahler moduli.
When the D-brane wraps a large cycle, this reduces to the above conditions. In particular, for a D7-brane with abelian flux  the field strength of the line bundle has to be of type $(1,1)$ and primitive, i.e.
\begin{equation}
\cF^{2,0}=0 \qquad\qquad  J\wedge \cF=0 \:.
\end{equation}
The second one implies that the flux generated FI-term vanishes.

When the cycle shrinks to zero size, the stability condition changes. For dP$_n$ singularities, i.e.\ point-like singularities arising when a dP$_n$ divisor in the compact manifold shrinks to zero size, the set of exceptional sheaves corresponding to stable fractional branes has been worked out \cite{Morrison:1996xf,Douglas:1996xp,Intriligator:1997pq,Hanany:2001py,Aspinwall:2004vm}. In this article, we are interested in fractional branes at dP$_0$ singularities. The corresponding sheaves have support on the shrinking dP$_0$ and are characterised by their Chern characters~\cite{Diaconescu:1999dt,Douglas:2000qw}\footnote{Note that we use the opposite sign convention with respect to the literature on D3-branes at dP$_n$ singularities. This is because in our convention a D7(anti-D7)-brane has charge $+1(-1)D$, where $D$ is the  wrapped divisor. Note again that in this convention, the D3-charge is minus the integral of the six-form component of $\Gamma_{D7}$.}
\bea\label{ChFi}
 \mbox{ch}({F_0}) = -1 + H -\tfrac12\, H\wedge H\,,  &
 \mbox{ch}({F_1}) = 2 - \,H -\tfrac12\, H\wedge H\,, &
 \mbox{ch}({F_2}) = - 1\,,
\label{fluxsheaf}
\eea
where $H$ is the hyperplane class of dP$_0=\mathbb{P}^2$.

\

In this paper, we study transitions between BPS D-brane configurations with the same total charge and, therefore, with the same total mass. Since these configurations are made up by several BPS objects, they satisfy the BPS condition only if the involved objects are mutually supersymmetric.
The sum of two BPS objects remains BPS if the phases of the central charges of the two objects are aligned.
Supersymmetric transitions take place among BPS configurations with the same charges and, therefore, the same mass/energy. Hence, these transitions occur at zero energy cost, i.e.\ are flat directions.

As we have said, the set of stable BPS objects can change if we vary the geometric (K\"ahler) moduli of the CY three-fold. Since we consider transitions among BPS states of the same set, we will keep these geometric moduli fixed and see what happens by varying the open string moduli.

In summary, if we have two different BPS configurations with the same charges, then there can be a (zero-energy cost) transition among them.\footnote{As a simple example consider the $\mathbb{C}_3/\mathbb{Z}_3$ singularity: we can move a `standard' D3-brane on top of the singularity. At this point in  moduli space the D3-brane splits into a set of three fractional branes. Both the D3-brane and the fractional branes at the singularity are BPS configurations with the same central charge. For a D3-brane we have $Z=-1$ while the central charges of the fractional branes are $Z_{\rm frac}=-1/3$.}

\subsection{Non-compact models}

\begin{figure}[t]\begin{center}
 \includegraphics[width=0.4\textwidth]{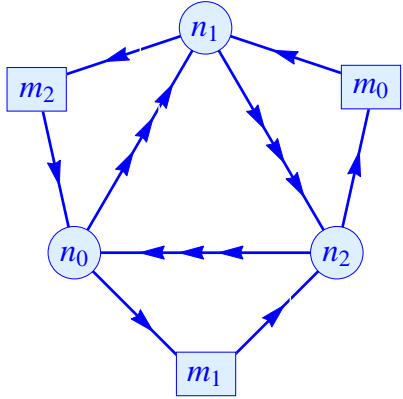}
\captionof{figure}{\footnotesize{The dP$_0$ quiver encoding the $SU(n_0)\times SU(n_1)\times SU(n_2)$ gauge theory with flavour branes. Potential D7-D7 states are not shown.}\label{fig:dp0quiverflav2}}
\end{center}\end{figure}

Let us consider the orbifold space $\mathbb{C}^3/\mathbb{Z}_3$. The singularity sits  at the fixed point $\{z_1,z_2,z_3\}=\{0,0,0\}\in\mathbb{C}^3$ of the $\mathbb Z_3$-action. If we place a D3-brane on top of this point, it splits into three fractional branes, with fractional central charge: $Z_{\rm frac}=\frac{Z_{D3}}{3}=-\frac13$. Consider a stack of $N$ D3-branes with gauge group $U(N)$. Once we put this stack on top of the singularity, the splitting produces three fractional branes with multiplicities $N$, a gauge group $U(N)^3$ and chiral bi-fundamental matter. One can also consider the case in which the multiplicities of the fractional branes differ. However, this requires (in order to cancel anomalies) the presence of D7-branes passing through the singularity and having non-zero chiral intersection with the fractional branes.
The resulting gauge theory can be represented  by a quiver diagram. For the $\mathbb{C}^3/\mathbb{Z}_3$ singularity, it is shown in Figure~\ref{fig:dp0quiverflav2}.
Each node with label $n_i$ corresponds to a distinct fractional brane stack with gauge group $U(n_i)$. The arrows indicate the bi-fundamental fields $(n_i,\bar{n}_j)$.  Given a choice of $n_i,$ the flavour D7 brane multiplicities\footnote{The numbers $m_i$ do not necessarily imply $U(m_i)$ gauge symmetries but can be, for instance, products of $U(1)$ gauge symmetries. Instead of one single arrow connecting the D3 and D7 branes, there may be multiple arrows with reduced gauge symmetry. All this is encoded in the choice of the $m_i$'s which are themselves determined by anomaly cancellation.}
$m_0$, $m_1$, $m_2$  are constrained by anomaly cancellation:
\begin{equation}\label{m0m2asm1ni}
 m_0 = m + 3(n_1-n_0)\ , \qquad m_1=m\ ,\qquad m_2 = m + 3(n_1-n_2)  \:.
\end{equation}

From a local point of view, models with different values of $n_i$ and $m$ are not related to each other, i.e.~there seems to be no flat direction in moduli space which permits a change of these numbers. In the following, we show that embedding these models in a globally consistent string compactification allows such transitions.

\subsection{Compact models}

We now want to embed the local models at the $\mathbb{C}^3/\mathbb{Z}_3$ singularity in a compact CY  manifold. Therefore, we have to consider a CY three-fold $X$ which admits such a singularity. This is the case if $X$ has a dP$_0$ divisor $\cD_{\textmd{dP}_0}$; in the limit in which this divisor shrinks to zero size, a $\mathbb{C}^3/\mathbb{Z}_3$ orbifold singularity is generated.

To embed the local D-brane model, we need the globally defined charge vectors of the fractional and flavour branes which produces the quiver diagram in Figure~\ref{fig:dp0quiverflav2}, with a chosen set of integers $n_i$ and $m$.
In \cite{Cicoli:2013mpa}, we have seen that this imposes strong constraints on the values of $n_i$ and $m$.

A fractional brane is a BPS brane with support on the shrinking dP$_0$. There are three types of mutually stable fractional branes for such a type of singularity. Their charge vectors are determined by \eqref{ChargeVectD7br}, with $D=\cD_{\textmd{dP}_0}$ (i.e.\ the shrinking divisor) and with ch$(\cE)$ given by~\eqref{ChFi}:
\begin{eqnarray}\label{FractionalD3ChVect}
 \Gamma_{F_0} &=& \cD_{\textmd{dP}_0}\wedge\left\{ -1  -\tfrac12 D_H  -\tfrac14  D_H \wedge D_H  \right\}\,, \nonumber\\
 \Gamma_{F_1} &=& \cD_{\textmd{dP}_0}\wedge\left\{ 2 +2  D_H + \tfrac12  D_H \wedge D_H  \right\}\,, \\
 \Gamma_{F_2} &=& \cD_{\textmd{dP}_0}\wedge\left\{ -1 - \tfrac32 D_H -\tfrac54  D_H \wedge D_H  \right\}\,.\nonumber
\end{eqnarray}
$D_H$ is a two-form of the CY three-fold whose pullback lies in the class $H$ of the dP$_0$ divisor.\footnote{There is an ambiguity in choosing $D_H$, as we can add to it any two-form of $X$ whose pullback onto the dP$_0$ is trivial.}

The flavour D7-branes are BPS branes wrapping a large divisor $\cD_{\rm flav}$ which pass through the singularity, i.e.\ in the resolved picture $\cD_{\rm flav}\cap \cD_{\textmd{dP}_0}\not =\emptyset$. Each of them will have an associated charge vector
\begin{equation}
 \Gamma_{\cE_{\rm flav},\cD_{\rm flav}} = \cD_{\rm flav}\wedge \mbox{ch} (\cE_{\rm flav}) \wedge \sqrt{\frac{\mbox{Td}(T\cD_{\rm flav})}{\mbox{Td}(N\cD_{\rm flav})}} \:,
\label{ChVectFlav}
\end{equation}
where $\cE_{\rm flav}$ is the vector bundle living on it. Since the flavour D7-brane extends in the non-compact directions in the local model, its global charge vector is not fully determined by the local data -- in contrast to the fractional brane.  The local ones, i.e.\ the pullback of \eqref{ChVectFlav} to $\cD_{\textmd{dP}_0}$, are given by:
\begin{equation}\label{LocalD7flCharg}
  \Gamma_{D7_i}^{\rm loc} \equiv \left.\Gamma_{\cE_{\rm flav}^i,\cD_{\rm flav}^i} \right|_{\textmd{dP}_0} = a_i H \wedge (1+ b_i H ) \qquad \mbox{with}\quad i=0,1,2\:,
\end{equation}
where the coefficient $a_i$ and $b_i$ are determined in \cite{Cicoli:2013mpa}:
\begin{equation}\label{LocFlavaibi}
  \{a_0,b_0\}=\{m_0 , \tfrac12\} \qquad  \{a_1,b_1\}=\{-2m_1 , 1\} \qquad  \{a_2,b_2\}=\{m_2 , \tfrac32\} \:,
\end{equation}
with $m_i$ as in~\eqref{m0m2asm1ni}.
Imposing that $ \Gamma_{D7_i}^{\rm loc} $ come from globally defined and connected flavour D7-branes gives the following constraints \cite{Cicoli:2013mpa}:
\begin{equation}\label{nogonmbis}
  0 \leq -m \leq 3(n_1 - \max\{n_0,n_2\})\,.
\end{equation}

\section{Transitions among different quiver gauge theories}
\label{sec:transition}

In the local picture, quiver models with different multiplicities $n_i$ of fractional branes are disconnected. As we now explain, this is not the case once one considers the full compactification with all the other branes needed to cancel the tadpoles. Due to bulk effects, there are smooth transition between two brane configurations related to different quiver gauge theories.

In the example studied in \cite{Cicoli:2013mpa}, we noticed that two different setups have the same D-brane charges. The first one was $n_1=3$, $n_0=n_2=2$ and two flavour branes; the second one was $n_0=n_1=n_2=3$ and no flavour brane but with two bulk branes that do not intersect the dP$_0$ divisor and rest identical with the first configuration. The same charges can also be realised by a configuration with $n_1=n_2=3$, $n_0=2$, one flavour brane and one bulk brane. All three configurations are made up by mutually BPS D-branes. Hence, we have evidence for a possible flat direction  connecting the three D-brane formations.
The basic step is the following: starting from one D7-brane that does not touch the singularity, we end up with one more flavour brane and one of the multiplicities  $n_i$ of the nodes changed. The transition occurs when the bulk brane touches the singularity.

Let us first consider a toy model for this process. We take the weighted projective space~$\mathbb{P}^3_{1,1,1,3}$. This toric space has the following weights and SR-ideal:
\begin{equation}
\begin{array}{|c|c|c|c|}
\hline x_0 & x_1 & x_2 & y   \tabularnewline \hline \hline
    1  &  1  &  1  &  3   \tabularnewline\hline
\end{array}  \qquad\qquad {\rm SR}=\{x_1\, x_2\, x_3\, y\}\,.
\end{equation}
This variety has one $\mathbb{C}^3/\mathbb{Z}_3$ singularity at $x_0=x_1=x_2=0$. Let us assume that we have three D3-branes on top of it. Now, take a D7-brane wrapping the holomorphic divisor
\begin{equation}
P_3(x_i)+\alpha\,y=0 \:.
\end{equation}
When $\alpha\not=0$ the singular point does not belong to the divisor. At $\alpha\rightarrow 0$, the D7-branes passes through the singularity: this is when the transition of the quiver system should occur. To understand better what is happening, we go to the resolved picture.

The resolved three-fold is given by
\vspace*{0.01\textwidth}
\begin{equation}
\begin{array}{|c|c|c|c|c|}
\hline x_0 & x_1 & x_2 & y & z \tabularnewline \hline \hline
    1  &  1  &  1  &  3  &  0 \tabularnewline\hline
    0  &  0  &  0  &  1  &  1 \tabularnewline\hline
\end{array}    \qquad\qquad {\rm SR}=\{x_1\, x_2\, x_3, y\,z\}\,.
\end{equation}
\vspace*{0.005\textwidth}

\begin{figure}[t]
\begin{center}
\includegraphics[width=0.5\textwidth]{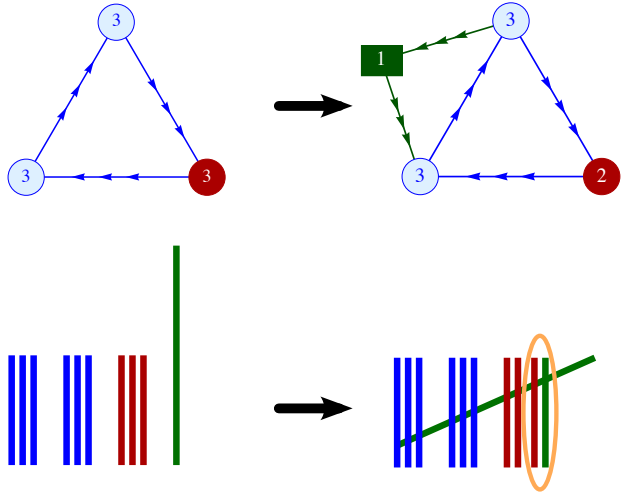}
\captionof{figure}{\footnotesize{Transition from the $SU(3)^3$ quiver to the $SU(3)^2\times SU(2)$: One bulk D7-brane (solid green line)  splits into a flavour brane intersecting the fractional branes (red and blue lines) and into an anti-fractional brane. This last one annihilates one fractional brane from the red set (yellow circle).
}}\label{fig:toytrans}
\end{center}
\end{figure}

\noindent
The blow-up divisor is the dP$_0$ given by $z=0$. In the resolved space, the proper transform of the equation for the bulk D7-brane is
\begin{equation}\label{InitalBulkRes}
z\cdot P_3(x_i)+\alpha\,y=0 \:.
\end{equation}
If we set now $\alpha\rightarrow 0$, the bulk D7-brane will split into a brane wrapping the dP$_0$ divisor at $z=0$ and a brane which intersect the blown-up dP$_0$. In the blown-down picture it is a brane passing through the singularity. From \eqref{InitalBulkRes} we see also that  the local D7-charge of the flavour brane equals $3H$. The brane wrapping the dP$_0$ will annihilate with one fractional brane either at the $n_0$-node or at the $n_2$-node, cf.~Figure~\ref{fig:dp0quiverflav2}. Correspondingly, the multiplicities of the associated fractional brane decreases. The change in the quiver system is shown in Figure~\ref{fig:toytrans}.
In the blown-down (physical) picture the transition will be between BPS configurations. What happens is that a bulk brane and a fractional brane `recombine'
to form a flavour brane
\be
D7_{\rm bulk}^{0,2} + D3_{\rm frac}^{0,2} \quad \leftrightarrow \quad  D7_{\rm flav}^{0,2}\,,
\ee
where the notation and the direction of the arrows have already been explained in Section~\ref{Introduction}.

\begin{figure}[t]
\begin{center}
\includegraphics[width=0.5\textwidth]{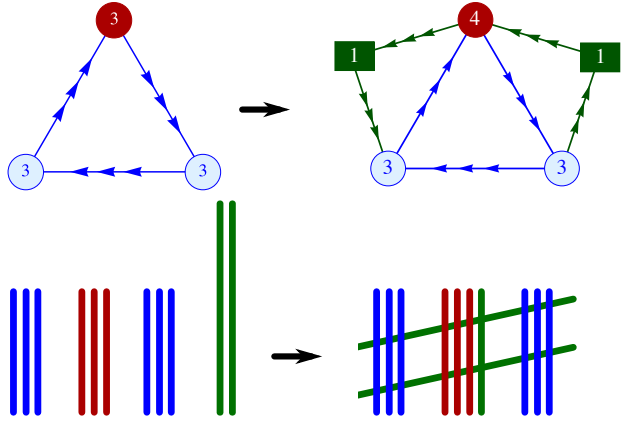}
\captionof{figure}{\footnotesize{Transition from the $SU(3)^3$ quiver to the $SU(4)\times SU(3)^2$ quiver: Two bulk D7-branes (solid green line)  splits into a flavour brane of type 0 and type 2 respectively, intersecting the fractional branes (red and blue lines) and into a fractional brane. }}\label{fig:dp04331}
\end{center}
\end{figure}

Let us consider a second type of transition.
We start from a stack of two D7-branes wrapping the divisor \eqref{InitalBulkRes}. When $\alpha\rightarrow 0$, the bulk stack splits into a rank-two brane passing through the singularity and a rank-two brane wrapping the dP$_0$ divisor.
The second stack can behave in two ways: either, as before, it annihilates with two fractional branes at the nodes $n_0$, $n_2$ or it increases the fractional brane stack at the node $n_1$. The two different final configurations have the same D7- and D5-charge, but different D3-charge. As we will see they differ by one unit of D3-charge. Hence, it will be the D3-charge of the initial bulk branes that determines which transition occurs.
The second transition can be summarised as follows (cf. Figure~\ref{fig:dp04331})
\be
2\,D7_{\rm bulk}^1  \quad \leftrightarrow \quad  D7_{\rm flav}^0 + D7_{\rm flav}^2 + D3_{\rm frac}^{1}\,,
\ee
where again  the notation and the direction of the arrows have been described in Section~\ref{Introduction}. Notice that the inverse of this transition corresponds to the recombination of two holomorphic branes in the resolved pictures.

\subsection{Kinematic conditions for the transitions}

Next, we study the kinematics of such transition in more detail. We will apply the following generic consideration to the explicit example in Section~\ref{sec:example}.

Let us consider a CY three-fold with one shrinking dP$_0$ four-cycle $\cD_{\textmd{dP}_0}$, or $\mathbb{C}^3/\mathbb{Z}_3$ singularity, plus its image under the orientifold involution. We start from a configuration of fractional branes and flavour branes that realise some dP$_0$ quiver gauge theory. We then take a bulk D7-brane wrapping a divisor $\cD$ that does not pass through the singularity, i.e.\ it does not intersect the shrinking dP$_0$ divisor in the resolved picture, $\cD\cap\cD_{\textmd{dP}_0}=\emptyset$.
We consider the deformation of the D7-brane which makes it split into a brane wrapping the dP$_0$ divisor and a (set of) flavour branes -- like the limit $\alpha\rightarrow 0$ in the toy model. The D7-brane wrapping the shrinking dP$_0$ will dissolve into a set of fractional (anti-)branes, changing the numbers $n_i$ of the dP$_0$ quiver theory; $m_i$ is modified by the additional flavour brane(s).
We now look under which conditions this D7-brane can split in this way.

First of all, the D-brane charges must be conserved during the splitting of the bulk brane, i.e.\ the sum of the charge vectors before and afterwards must agree:\footnote{We neglect the contribution coming from the $B$-field, as it gives only a common factor $e^{-B}$ to the charge vectors.}
\be
\label{GenChConsTr}
\Gamma_{\cD} = \Delta\Gamma_{\rm flav} + \Delta\Gamma_{\textmd{dP}_0}\,,
\ee
where $\Gamma_{\cD}$ is the charge vector \eqref{ChargeVectD7br} of the bulk brane wrapping the divisor $\cD$. $\Delta\Gamma_{\rm flav}$  is the charge vector of the additional flavour branes:
\be
\label{ChVectDeltaflavour}
\Delta\Gamma_{\rm flav}=
\Gamma_{D7^{\rm flav}}^0 + \Gamma_{D7^{\rm flav}}^1 + \Gamma_{D7^{\rm flav}}^2  \:,
\ee
with $\Delta\Gamma_{\rm flav}$ the sum of the produced flavour branes and $\Gamma_{D7^{\rm flav}}^i$ are the charge vectors of the three kinds of local flavour branes given in~\eqref{LocalD7flCharg}. 
$\Delta\Gamma_{\textmd{dP}_0}$ is the shift in the fractional branes
\be
\label{ChVectDeltadP0}
\Delta\Gamma_{\textmd{dP}_0} =
\cD_{\textmd{dP}_0}\wedge \mbox{ch}(\cE) \wedge \sqrt{\frac{\mbox{Td}(T\cD_{\textmd{dP}_0})}{\mbox{Td}(N\cD_{\textmd{dP}_0})}} \qquad\mbox{with }\, \mbox{ch}(\cE)=r + n H +(\tfrac{1}{2} n^2-\ell)H^2\,.
\ee

Let us expand the condition \eqref{GenChConsTr}.
The D7-charge conservation implies the following condition on the divisors wrapped by the branes:
\begin{equation}\label{D7chcons}
 \cD = r\cD_{\textmd{dP}_0} + \sum_{i=0,1,2}  \cD_{\rm flav}^{(i)}\:.
\end{equation}
In order for the described transition to occur, we need that all the involved divisors classes have holomorphic connected representatives -- otherwise we would have a different splitting. For instance, if there is no holomorphic smoothly connected divisor in the class $\cD_{\rm flav}=\cD-\cD_{\textmd{dP}_0}$, we cannot have the transition described in the toy model.

The D5-charge conservation constrains the gauge bundle:
\begin{equation}\label{D5chcons}
  \cD\wedge \textmd{tr} F = \cD_{\textmd{dP}_0}\wedge \left(n+\tfrac32\right)H + \sum_{i=0,1,2} \cD_{\rm flav}^{(i)}\wedge \textmd{tr} F_{\rm flav}^{(i)}\:.
\end{equation}
The fluxes $\textmd{tr} F_{\rm flav}$ on the new flavour branes can be written as
\begin{equation}\label{Flflsplit}
  \textmd{tr} F_{\rm flav}^{(i)} = \tilde{F}_{\rm flav}^{(i)} + \beta_{\textmd{dP}_0}^{(i)} \cD_{\textmd{dP}_0}\:,   \qquad\mbox{with}\qquad
  \beta_{\textmd{dP}_0}^{(i)}\in\{-\tfrac16,-\tfrac13,-\tfrac12\}\:,
\end{equation}
and $\tilde{F}_{\rm flav}$ being a two-form orthogonal to $\cD_{\textmd{dP}_0}$ on the D-brane world-volume. The coefficient $\beta_{\textmd{dP}_0}$ is determined by the local D5-charge of the fractional branes~\eqref{LocFlavaibi}.
Moreover, if $\Delta m_i$ are the shifts of the flavour branes, their restriction on the dP$_0$ is equal to $\cD_{\rm flav}^{(i)}|_{\textmd{dP}_0} = ( \Delta m_0\,H, -2\Delta m_1\,H,\,\Delta m_2\,H)_i$.
Hence, we can rewrite the D5-charge conservation condition as
\begin{equation}\label{D5chConsGenpre}
    \cD \wedge \textmd{tr} F =  \sum_{i=0,1,2}\cD_{\rm flav}^{(i)}\wedge \tilde{F}_{\rm flav}^{(i)}+ \cD_{\textmd{dP}_0}\wedge \left(  n+\tfrac32 -\tfrac16 \Delta m_0 + \tfrac23 \Delta m_1 - \tfrac12 \Delta m_2 \right)H\:. \nonumber
\end{equation}
We will see in a moment, equation~\eqref{Deltami}, that the second term on the right hand side vanishes, leading to the simple form:
\begin{equation}\label{D5chConsGen}
    \cD \wedge \textmd{tr} F =  \sum_{i=0,1,2}\cD_{\rm flav}^{(i)}\wedge \tilde{F}_{\rm flav}^{(i)} \:.
\end{equation}

Finally one needs to check that the D3-charge is conserved. Given a certain gauge bundle on the bulk brane wrapping $\cD$ and requiring, for example, to have only abelian fluxes on the flavour branes,  the coefficient~$\ell$ in~\eqref{ChVectDeltadP0} is determined.

\

So far we have analysed the condition for the bulk brane to split into a brane wrapping the dP$_0$ and a set of branes intersecting it -- the flavour branes.
After the splitting, the D7-brane wrapping the dP$_0$ has a charge vector equal to \eqref{ChVectDeltadP0}. For generic $r$, $n$ and $\ell$, this brane is not a fractional brane but a combination of them. However, the only stable branes on the shrinking dP$_0$ are the fractional branes. Therefore the brane must dissolve into them:
\begin{equation} \label{ChSplitFi}
 \mbox{ch}(\cE) = \Delta n_0 \, \mbox{ch}(F_0) + \Delta n_1 \, \mbox{ch}(F_1) + \Delta n_2 \, \mbox{ch}(F_2) \:,
\end{equation}
where the Chern characters ch$(F_i)$ of the fractional branes were given in \eqref{ChFi}. The multiplicities of the fractional branes after the transition are $n_i'=n_i+\Delta n_i$ with $i=0,1,2$. Imposing the equality of \eqref{ChSplitFi} with the expression of ch$(\cE)$ in \eqref{ChVectDeltadP0}, one obtains
\begin{equation}
  \Delta n_0 = -\tfrac12 n(n-1) + \ell \qquad  \Delta n_1 = -\tfrac12 n(n+1)+ \ell \qquad  \Delta n_2 = -\tfrac12 n(n+3) -r+ \ell\:.
\end{equation}
Notice that, since $r>0$, for $n\in\mathbb{Z}$ we have $\Delta n_i -\ell\leq 0$ $\forall i$.
A negative $\Delta n_i$ means that the splitting generates anti-fractional branes which annihilates $-\Delta n_i$ fractional branes of  type~$i$ in the initial quiver system. We also see that the $\Delta n_i$ are all shifted by $\ell$. Increasing $\ell$ means to increase the D3-charge
without modifying its D5- and D7-charges, cf.\ ch$(\cE)$ in \eqref{ChVectDeltadP0}. Hence, the transitions with equal $\Delta n_i$, up to an integer overall shift $\ell$, differ by absorption or ejection of $|\ell|$ D3-brane.

The splitting will also produce a number of new flavour branes, whose multiplicities are given by
\begin{equation}\label{Deltami}
 \Delta m_0 = \Delta m -3n \qquad  \Delta m_1 = \Delta m \qquad  \Delta m_2 = \Delta m + 3(n+r)\:,
\end{equation}
where $\Delta m$ signals the arbitrariness in choosing which flavour branes are generated in the transition. Since the restrictions of the flavour brane divisors on  dP$_0$ are:
\begin{equation}\label{DflavRestrdP}
 \cD_{\rm flav}^{(0)} |_{\textmd{dP}_0}= \Delta m_0\,H \qquad \cD_{\rm flav}^{(1)} |_{\textmd{dP}_0}= -2\Delta m_1\,H \qquad \cD_{\rm flav}^{(2)} |_{\textmd{dP}_0}= \Delta m_2\,H\:,
\end{equation}
the following constraints on the values of $\Delta m_i$ are imposed from the global embedding \cite{Cicoli:2013mpa}:
\begin{equation}\label{DmiConstr}
 \Delta m \leq 0 \qquad\qquad -r-\tfrac13 \Delta m \leq n \leq \tfrac13\Delta m \:.
\end{equation}
We are interested in two cases which can be used to connect all quiver models:
\begin{enumerate}
\item $r=1$: from  \eqref{DmiConstr}, we must have $\Delta m =0$ and $n=0,-1$. If we also take $\ell=0$, we obtain the transition described in the above toy model, where the bulk brane splits into a flavour and an anti-fractional brane of type $0$ or type $2$. The anti-brane annihilates the corresponding fractional brane and, in fact, we have either $\Delta n_0=-1$ or $\Delta n_2=-1$. By repeating such transitions, we can lower the multiplicities $n_0$ and $n_2$ and simultaneously generating the necessary additional flavour branes.
\item $r=2$: in this case $-3\leq \Delta m\leq 0$. We will be interested in $\Delta m=0$ and $n=-1$. If we choose also $\ell=1$, we have $\Delta n_1=1$, $\Delta n_0=\Delta n_2=0$, $\Delta m_1=0$ and $\Delta m_0=\Delta m_2=1$. Here the bulk D7-brane splits into the fractional brane at the $n_1$ node ($\Delta n_1=+1$) and into two flavour branes. In this case we have no annihilation.
\end{enumerate}
We consider the case $\Delta m=0$ because the other possible values of $\Delta m$ correspond to a recombination
of a flavour brane of type $0$ with a flavour brane of type $2$ which gives a flavour brane of type $1$. Of course, whether this recombination is possible depends on the gauge bundles on the two flavour branes.

We will illustrate these two classes of transitions in Section~\ref{sec:example}.

\subsection{F- and D-flatness conditions}

We consider transitions in which both the initial and the final states are supersymmetric configurations.
The F- and D-term vanishing conditions are realised by requiring the stability conditions described in Section~\ref{sec:BPS_D-branes}
(if the VEVs of the charged fields are zero).

The system of the fractional branes with Chern characters \eqref{ChFi} is supersymmetric at the singular point \cite{Douglas:2000gi,Douglas:2000qw}. The flavour branes and the bulk branes must wrap holomorphic divisors $D$ and have holomorphic field strength, i.e.\ $\cF \in H^{1,1}(D)$, where the generated FI-terms vanish.

We will mostly consider D7-branes with abelian fluxes. In this case, the holomorphicity condition on $\cF$ is automatically satisfied if we take $\cF$ to be a two-form pulled back from the CY three-fold. When this is not the case, the holomorphicity condition may fix some D7-brane deformation moduli. The transition can occur if the corresponding flat direction is not lifted by the flux.

As regarding the D-terms, after taking the limit of shrinking dP$_0$ ($\tau_{\cD_{\textmd{dP}_0}}\rightarrow 0$), the FI-terms of the fractional branes vanish. Furthermore, the sum of the FI-terms of the flavour branes generated after the transition becomes equal to the FI-term of the initial bulk brane:
\begin{equation}\label{xiGenCase}
  \xi_{\cD_{\rm flav}^{(0)}}+\xi_{\cD_{\rm flav}^{(1)}}+\xi_{\cD_{\rm flav}^{(2)}} \qquad \underset{\tau_{\cD_{\textmd{dP}_0}}\rightarrow 0}{\longrightarrow}\qquad \xi_{\cD} = \frac{1}{\cV} \int_{\cD} \cF \wedge J \:,
\end{equation}
where we used the D5-charge conservation condition.
Consider the case when only one new flavour brane is generated: if the starting bulk brane has zero FI-term, i.e.\ $\xi_{\cD} = 0$, then the flavour brane will also have a vanishing FI-term.

\section{Transitions in an explicit dP$_0$ example}
\label{sec:example}

In this section we will apply the above considerations to an explicit example where
we will be able to describe in detail the transitions among different quiver gauge theories.
We consider the CY three-fold $X$ described in detail in~\cite{Cicoli:2012vw}. The hypersurface CY has the Hodge numbers  $h^{1,1}=4$ and $h^{1,2}=112$. Its ambient space is defined by the following weight matrix and Stanley-Reisner ideal:
\begin{equation}
\begin{array}{|c|c|c|c|c|c|c|c||c|}
\hline z_1 & z_2 & z_3 & z_4 & z_5 & z_6 & z_7 & z_8 & D_{eq_X} \tabularnewline \hline \hline
    1  &  1  &  1  &  0  &  3  &  3  &  0  &  0  & 9\tabularnewline\hline
    0  &  0  &  0  &  1  &  0  &  1  &  0  &  0  & 2\tabularnewline\hline
    0  &  0  &  0  &  0  &  1  &  1  &  0  &  1  & 3\tabularnewline\hline
    0  &  0  &  0  &  0  &  1  &  0  &  1  &  0  & 2\tabularnewline\hline
\end{array}
\label{eq:model3dP0:weightm}\, .
\end{equation}
\begin{equation}
{\rm SR}=\{z_4\, z_6,\,z_4\, z_7, \, z_5\, z_7,\, z_5\, z_8,\, z_6\, z_8,\, z_1\, z_2\, z_3\}\,.\nonumber
\end{equation}
The last column in~\eqref{eq:model3dP0:weightm} refers to the degrees of the hypersurface equation $eq_X=0$, with $eq_X$ given by
\begin{eqnarray}
\label{EqCY3}
 eq_X &\equiv& P^1_{3}(z_4 z_5-z_6 z_7)^2z_8 + P^2_{3}(z_4 z_5+z_6 z_7)^2z_8
      \nonumber \\ &&+ (P_{0}^+ z_5 z_6 + P^+_{6} z_4 z_7 z_8^2)(z_4 z_5+z_6 z_7) + P^+_{9} z_4^2 z_7^2 z_8^3 \\
	&& + (P_{0}^- z_5 z_6 + P^-_{6} z_4 z_7 z_8^2)(z_4 z_5-z_6 z_7)\,,\nonumber
\end{eqnarray}
where $P^{\pm,1,2}_{k}$ are polynomials in the coordinates $(z_1,z_2,z_3)$ of degree $k$. There is a basis of $H^{1,1}(X)$ such that the intersection form simplifies considerably:\footnote{Note that this basis of integral cycles is not an `integral basis'; in particular $D_1=\frac{1}{3}(\cD_b-\cD_{q_1}-\cD_{q_2}-\cD_s)$.}
\be
\cD_b = D_4+D_5=D_6+D_7,\qquad \cD_{q_1} = D_4,\qquad \cD_{q_2} = D_7,\qquad \cD_s = D_8\,,
\label{eq:simplebasis}
\ee
with
\be
\label{intersnumb}
I_3 = 27\,\cD_b^3 + 9\,\cD_{q_1}^3 + 9\,\cD_{q_2}^3 + 9\,\cD_s^3 \:.
\ee
The three elements $\cD_{q_1},\cD_{q_2},\cD_s$ are all dP$_0$ divisors on the CY. Moreover, the first two are exchanged by the involution
\be
\label{eq:OrInvol}
 z_4\leftrightarrow z_7\qquad\textmd{and}\qquad z_5\leftrightarrow z_6\:.
\ee
The orientifold-planes associated with this involution are $O7_1$ at $z_4z_5-z_6z_7=0$ (wrapping $\cD_b$) and $O7_2$ at $z_8=0$ (wrapping $\cD_s$) \cite{Cicoli:2012vw}. Moreover the equation of the symmetric CY is now given by \eqref{EqCY3} with $P_k^-\equiv0$.
In the following, we will consider  dP$_0$ quiver gauge theories constructed `around' the shrinking $\cD_{q_1}$ divisor at $z_4=0$ and its orientifold image $\cD_{q_2}$ at $z_7=0$.

Let us consider the trinification $SU(3)^3$ model described in \cite{Cicoli:2012vw} with four bulk D7-branes plus their images wrapping the divisor $\cD_b$ to cancel the D7-tadpole generated by the O7-plane $O7_1$. We will take the abelian flux $F_b$ on the bulk branes to be of type $(1,1)$, i.e. $F_b\in H^{1,1}(\cD_b)$, and to have zero FI-term.
The FI-term depends on the invariant combination $\cF_b =F_b-B$: $\xi\propto \int_{\cD_b}J\wedge \cF_b$. This is zero
if the flux $\cF_b$ is the Poincar\'e dual of a two-cycle of $\cD_b$ which is trivial in the CY three-fold, even though not necessarily trivial on $\cD_b$.  In this case, however,  we might need to fix some D7-brane deformations to make the two-form holomorphic. Moreover, $\cF_b$ satisfy this condition only if the part of $F_b$ proportional to $\frac{c_1(\cD_b)}{2}=-\frac{\cD_b}{2}$ (necessary to prevent a Freed-Witten anomaly~\cite{Minasian:1997mm,Freed:1999vc}) is cancelled by the B-field. For this reason we take $B=-\frac{\cD_b}{2}-\frac{\cD_s}{2}$.\footnote{The second term is present to cancel the Freed-Witten flux on the non-perturbative cycle, but it is irrelevant for this discussion.}
In principle one could obtain vanishing FI-terms by fixing some combinations of K\"ahler moduli \cite{Cicoli:2011qg}; in the present case this can be shown to be impossible, due to the simple intersection form \eqref{intersnumb} which in the literature has been called `strong Swiss-cheese' \cite{Cicoli:2011it} .

\subsection{Transitions of type I}
\begin{figure}[t]\begin{center}
\includegraphics[width=0.5\textwidth]{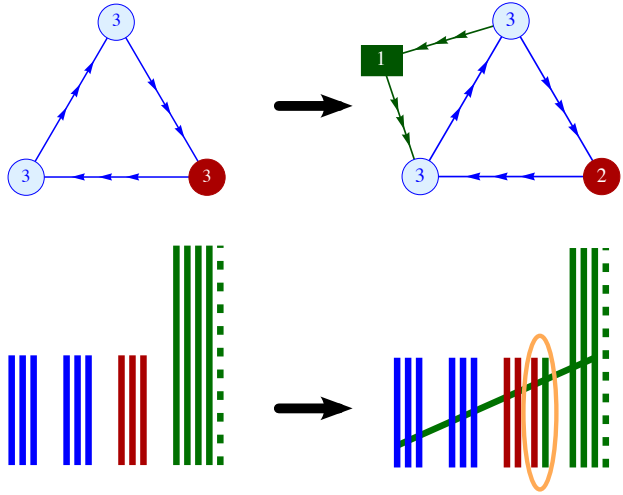}
\captionof{figure}{\footnotesize{Transition from the $SU(3)^3$ quiver to the $SU(3)^2\times SU(2)$: One D7-brane (solid green line) on top of the O-plane (dotted line) splits into a flavour brane intersecting the fractional branes (red and blue lines) and an anti-fractional brane which annihilates with a fractional brane from the red set (yellow circle).
}\label{transSU3SU2}}
\end{center}\end{figure}
Let us start with the simple case of a bulk D7-brane wrapping a divisor in the class $\cD_b$. The most generic equation describing such a divisor is
\begin{equation}\label{DbDiv}
 z_4z_5 + \alpha z_6z_7 + P_3^{D7}(z_1,z_2,z_3)\,z_4z_7z_8\:,
\end{equation}
with $P_3^{D7}(z_1,z_2,z_3)$ a homogeneous polynomial of degree three in the coordinates $z_1$, $z_2$ and $z_3$.
For $\alpha=-1$ and $P_3^{D7}\equiv 0$, the bulk brane is on top of the orientifold plane $O7_1$.
When $\alpha\rightarrow 0$ in eq.~\eqref{DbDiv}, the bulk brane touches the singularity at $z_4=0$. In the resolved picture, the bulk brane equation becomes
\begin{equation}
 z_4 \cdot (z_5 + P_3^{D7}(z_1,z_2,z_3)\,z_7z_8)\,.
\end{equation}
It splits into one D-brane wrapping the divisor $\cD_{q_1}$ and one D-brane wrapping the divisor $D_5=\cD_b-\cD_{q_1}$.
This second brane has the D7-charge of a flavour brane. In fact, $D_5|_{\cD_{q_1}}=3H$. Hence, it is a flavour brane with either $m_0=3$ or $m_2=3$. The D-brane wrapping the dP$_0$ will annihilate either a brane at the node $n_0$ or one at the node $n_2$ (cf. Figure~\ref{transSU3SU2}). It is the flux $\cF_b$ that determines which possibility is realised.

Notice that in this transition the number of chiral massless fermions remains unchanged, even is these modes get rearranged in different representations of the new gauge group.

\subsubsection{Transitions of type I with $\Delta n_2=-1$}

Let us make the most simple choice, i.e.\ $\cF_b=0$, which means $F_b=-\frac{\cD_b}{2}$ as $B=-\frac{\cD_b}{2}-\frac{\cD_s}{2}$; the quantisation of $F_b$ is consistent with the cancellation of the Freed-Witten anomaly. The charge vector of the bulk brane is then
\begin{equation}\label{ChVectDbZeroFlux}
  \Gamma_{\cD_b} = \cD_b + \frac{39}{8} d\textmd{Vol}_X^0 \:.
\end{equation}
The D-brane charges are conserved by the following splitting:
\begin{equation}
  \Gamma_{\cD_b} \,\, \rightarrow \,\,
  \left\{\begin{array}{ccl}
  \Gamma_{\cD_{\textmd{dP}_0}}&=& \cD_{q_1} + \cD_{q_1}\wedge \tfrac32 D_1 + \tfrac54 d\textmd{Vol}_X^0 =-\Gamma_{F_2}\\
  \Gamma_{\cD_{\rm flav}^{(2)}}&=& (\cD_b-\cD_{q_1}) + (\cD_b-\cD_{q_1})\wedge (\tfrac32 D_1-\tfrac12\cD_b) + \tfrac{29}{8} d\textmd{Vol}_X^0 \\
  \end{array}\right. \:.
\end{equation}
The brane wrapping the dP$_0$ has minus the charge vector of a type $2$ fractional brane, i.e.\ an anti-fractional brane.
The second brane is a flavour brane with $\cD_{\rm flav}=(\cD_b-\cD_{q_1})$ and $\cF_{\rm flav}=\tfrac32 D_1-\tfrac12\cD_b$.
It is a flavour brane of type $2$, as $\cD_{\rm flav}|_{\cD_{q_1}}=3H$ and $\cF_{\rm flav} |_{\cD_{q_1}}=\frac32 H$.
This is then a transition with $\Delta n_2=-1$ and $\Delta m=0$.
One can check that, even if the flux on the flavour brane is non-trivial, its FI-term vanishes when the dP$_0$ is shrunk to zero size, as shown in~\eqref{xiGenCase}.

\subsubsection{Transitions of type I with $\Delta n_0=-1$}

To obtain a transition with $\Delta n_0=-1$, we need to choose a different flux on the bulk brane.
Since we still want a zero FI-term $\xi_b=0$, the flux must be of the non-pullback type. If the flux would be a pullback two-form, we would get $\xi_b\propto \cV^{1/3}$ which in turn would force the CY to collapse.

In appendix~\ref{AppFlux}, we obtain such a  flux $\cF_b$ by defining a non-pullback two-form $F^{\mathcal C}$ on $\cD_b$. After including the contribution from the $B$-field ($B=- \cD_b/2 - \cD_s/2$), we have:
\begin{equation}\label{trivialFlDb}
 \cF_b = \omega^\cC - \iota_\ast D_1   \:.
\end{equation}
The two-form $\cF_b$ is non-trivial on $\cD_b$. On the other hand, the push-forward of the Poincar\'e dual curve is trivial inside the CY $X$. This means that we, again, have vanishing D5-charge for the bulk brane and, therefore, zero FI-term. In appendix~\ref{AppFlux}, we also compute the square of $\cF_b$:
$$\int_{\cD_b}\cF_b^2=-12.$$
This allows us to write down the charge vector of the bulk brane:
\begin{equation}\label{ChVectDbtrivialFlux}
  \Gamma_{\cD_b} = \cD_b - \frac{9}{8} d\textmd{Vol}_X^0 \:.
\end{equation}
For the following transition the charges are conserved:
\begin{equation}
  \Gamma_{\cD_b} \,\, \rightarrow \,\,
  \left\{\begin{array}{ccl}
  \Gamma_{\cD_{\textmd{dP}_0}}&=& \cD_{q_1} + \cD_{q_1}\wedge \tfrac12 D_1 + \tfrac14 d\textmd{Vol}_X^0  = - \Gamma_{F_0}\\
  \Gamma_{\cD_{\rm flav}^{(0)}}&=& (\cD_b-\cD_{q_1}) + (\cD_b-\cD_{q_1})\wedge (\tfrac12 D_1+ \omega_{\rm flav}^{\cC} - \tfrac12\cD_b) - \tfrac{11}{8} d\textmd{Vol}_X^0 \\
  \end{array}\right. \:.
\end{equation}
The brane wrapping the dP$_0$ has minus the charge vector of a type $0$ fractional brane, i.e.\ an anti-fractional brane. The second brane is a flavour brane with $\cD_{\rm flav}=\cD_b-\cD_{q_1}$ and $\cF_{\rm flav}=\omega_{\rm flav}^{\cC} +\tfrac12 \iota_\ast D_1 - \iota_\ast B$. Here $\omega_{\rm flav}^{\cC}$ is again a non-pullback two-form defined by the curve $\cC$, as explained in appendix~\ref{AppFlux}.
Since $\cD_{\rm flav}|_{\cD_{q_1}}=3H$ and $\cF_{\rm flav} |_{\cD_{q_1}}=\frac12 H$, the flavour brane is of type~$0.$ This last result depends on the fact that the defining curve $\cC$ does not intersect the dP$_0$ divisor, and therefore $\omega^\cC_{\rm flav}|_{\cD_{q_1}}=0$.

In this transition, we have a different feature with respect to the previous one. This is due to the presence of a flux that is not of the pullback type. Such flux is not necessarily `automatically' holomorphic, and hence one may need to fix D7-brane deformations to keep the two-form of $(1,1)$-type. As explained in appendix~\ref{AppFlux}, switching on the flux \eqref{trivialFlDb} on \eqref{DbDiv} can fix the deformation parameter encoded in the coefficient of the polynomial $P^{D7}_3(z_1,z_2,z_3)$. In other words, varying $P^{D7}_3$ can make the curve $\cC$ non-holomorphic and break supersymmetry. Regardless of this issue, $\alpha$ remains free to vary in a supersymmetric transition. This deformation is the one that we need to realise the transition to the flavour brane wrapping $z_5+P^{D7}_3(z_1,z_2,z_3)\,z_7 z_8=0$. Note that the flux $\cF_{\rm flav}$ present on the flavour brane is constructed by using the curve $\cC$ which still fixes the deformations in $P^{D7}_3(z_1,z_2,z_3)$.

\subsection{Transitions of type II}
\begin{figure}[t]
\begin{center}
\includegraphics[width=0.5\textwidth]{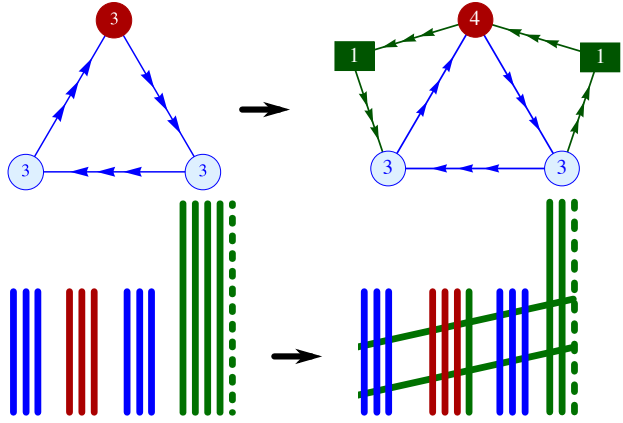}
\captionof{figure}{\footnotesize{Transition from the $SU(3)^3$ quiver to the $SU(4)\times SU(3)^2$ quiver: A rank-two bulk D7-brane (solid green line) 
splits into a fractional brane and two flavour branes (of type $0$ and $2$ respectively) intersecting the fractional branes (red and blue lines).
}\label{fig:dp0433}}
\end{center}\end{figure}
We start from a stack of two bulk D7-brane wrapping the divisor defined by equation \eqref{DbDiv}. This time, we can have a non-abelian flux, i.e.\ a non-trivial $SU(2)$-bundle. We still insist that the first Chern class of the bundle
has zero push-forward on the CY $X$, in order to have a zero FI-term. On the other hand, we allow for a non-trivial second Chern class $c_2$. In particular, we choose a brane wrapping $\cD_b$ with a rank two vector bundle characterised by the following Chern character
\begin{equation}\label{chEbulk}
  \mbox{ch} (\cE_{\rm bulk})= 2 + c_1(\cE_{\rm bulk}) + \left( \tfrac12 c_1(\cE_{\rm bulk})^2-c_2(\cE_{\rm bulk})\right) \:,
\end{equation}
with $c_1(\cE_{\rm bulk})=\omega^\cC - \iota_\ast D_1$
and $c_2=1\cdot d\textmd{Vol}_{\cD_b}^0$. Plugging this into \eqref{ChargeVectD7br}, we obtain the charge vector
\begin{equation}
  \Gamma_{\cD_b} = 2\cD_b + \frac{11}{4} d\textmd{Vol}_X^0 \:.
\end{equation}
A charge conserving transition is given by:
\begin{equation}
  \Gamma_{\cD_b} \,\, \rightarrow \,\,
  \left\{\begin{array}{ccl}
  \Gamma_{\cD_{\textmd{dP}_0}}&=& 2\cD_{q_1} + 2\cD_{q_1}\wedge D_1 + \tfrac12 d\textmd{Vol}_X^0  = +\Gamma_{F_1}\\
    \Gamma_{\cD_{\rm flav}^{(0)}}&=& (\cD_b-\cD_{q_1}) + (\cD_b-\cD_{q_1})\wedge (\tfrac12 D_1+ \omega_{\rm flav}^{\cC} - \tfrac12\cD_b) - \tfrac{11}{8} d\textmd{Vol}_X^0 \\
  \Gamma_{\cD_{\rm flav}^{(2)}}&=& (\cD_b-\cD_{q_1}) + (\cD_b-\cD_{q_1})\wedge (\tfrac32 D_1-\tfrac12\cD_b) + \tfrac{29}{8} d\textmd{Vol}_X^0 \\
    \end{array}\right. \:,
  \end{equation}
We see that we generate both kinds of flavour branes which we obtained in the two type I transitions. The D-brane wrapping the dP$_0$ divisor is a fractional brane of type $1$ and will increase the multiplicity $n_1$ in the quiver diagram (see Figure~\ref{fig:dp0433}).

\subsection{Step by step transitions}

We can now use the transitions described above to connect different quiver gauge theories. Take, for instance, the trinification model $SU(3)^3$ and apply one transition of type I with $\Delta n_0=-1$ and one of type I with $\Delta n_2=-1$. In this way we reach the left-right symmetric model with gauge group $SU(3)\times  SU(2)^2$. In this process two bulk D7-branes wrapping the divisor $\cD_b$ are transformed into the two needed flavour branes. Moreover, if one consider also the D7-D7 states (between the two flavour branes), the total number of chiral modes remains the same as in the $SU(3)^3$ model (even if they are rearranged in different representations of the new gauge group).\footnote{This happens in the explicit example we have considered, where the only contribution to the flavour brane fluxes comes from the local flux necessary to cancel the local D5-charge.}

\begin{figure}[t]\begin{center}
\includegraphics[width=0.17\textwidth]{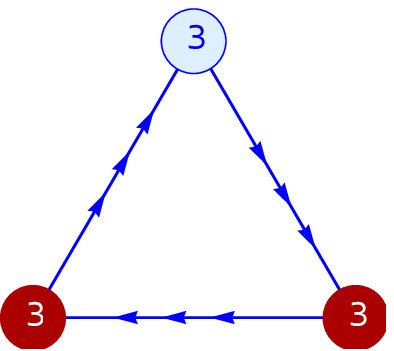}\hspace{0.4cm}\includegraphics[width=0.10\textwidth]{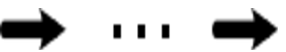}\includegraphics[width=0.23\textwidth]{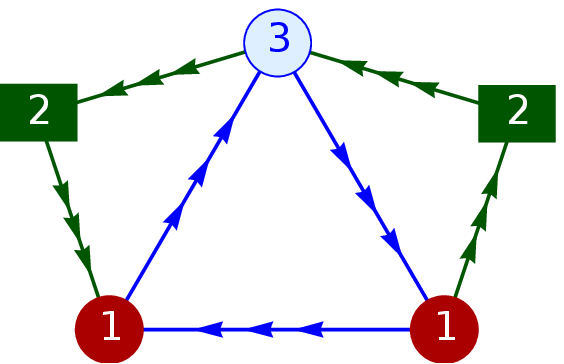}\\
\includegraphics[width=0.52\textwidth]{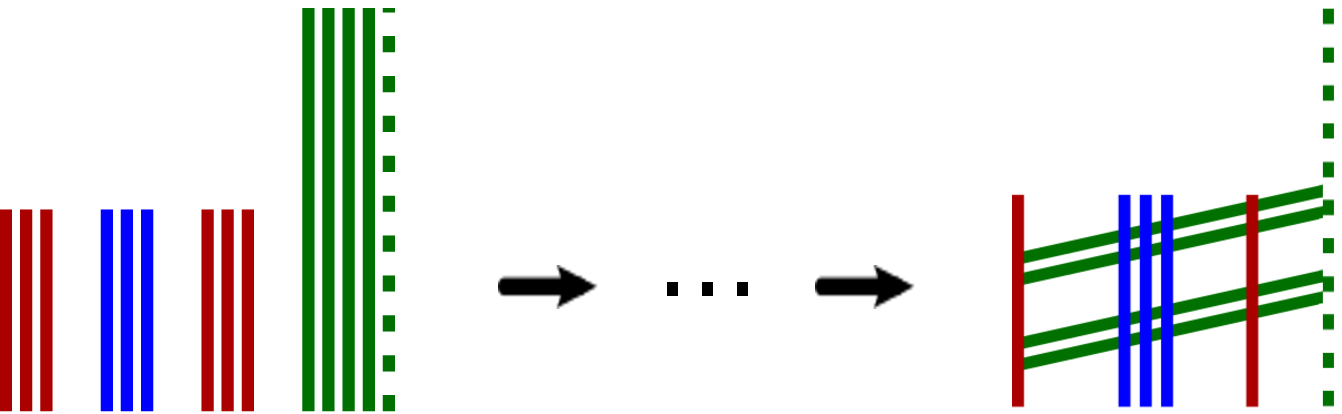}
\captionof{figure}{\footnotesize{By repeating four times the transition described in Figure~\ref{transSU3SU2}, we get four flavour branes and no D7-brane on top of the O-plane.}\label{TransSU3SU1}}
\end{center}\end{figure}

One can do a step forward and apply another transition of type I.
The final gauge group on the fractional branes is the Standard Model, i.e.\ $SU(3)\times SU(2)\times U(1)$, with the proper structure of flavour branes.
In our case one can, in principle, go one step further: on top of the O7-plane wrapping $\cD_b$ there is still one more D7-brane. If this undergoes the transition as well, either $n_0$ or $n_2$ are lowered by one,
reaching the situation depicted in Figure~\ref{TransSU3SU1}.

\begin{figure}[t]\begin{center}
\includegraphics[width=0.8\textwidth]{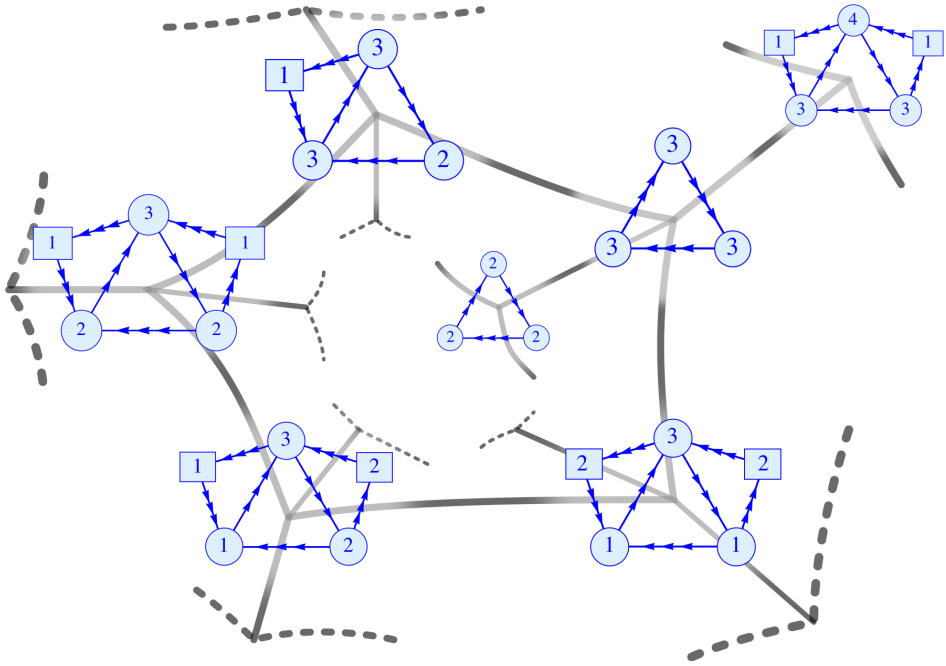}
\captionof{figure}{\footnotesize{Part of the web of quiver gauge theories at a dP$_0$ singularity. In particular we show the gauge theories which have been shown to be in direct connection with the trinification $SU(3)^3$ model. Different levels in the web correspond to different D3-brane charges. Only some connections in the web are shown.}\label{fig:theweb}}
\end{center}\end{figure}

Starting from the $SU(3)\times SU(2)^2$ model we can take another route. We may apply the reverse of the type II transition: the two flavour branes recombine with the fractional brane of type $1$:
\begin{equation}
  \left.\begin{array}{lcl}
 \Gamma_{F_1}\\
  \Gamma_{\cD_{\rm flav}^{(0)}}\\
  \Gamma_{\cD_{\rm flav}^{(2)}}\\
  \end{array}\right\} \,\,\rightarrow \,\, \Gamma_{\cD}^{\rm n.a.}= 2\cD_b  + \tfrac{11}{4}d\textmd{Vol}_X^0 = \cD_b \wedge  \mbox{ch}(\cE_{\rm bulk})\wedge  \sqrt{\frac{T\cD_b}{N\cD_b}}\:,
\end{equation}
where $ \mbox{ch}(\cE_{\rm bulk})$ is given in \eqref{chEbulk}.
We see that we get a bulk brane wrapping the divisor $\cD_b$, with a bundle $\cE_{\rm bulk}$ of rank two and second Chern class such that $\int_{\cD_b}c_2(V)=1$, i.e.\ we have a brane-stack with an instanton-like background in the extra-dimension of $\cD_b$ with instanton number equal to $1$. The moduli space of the instanton-like solution has a flat direction that is the size of the instanton. At the point in moduli space where the instanton becomes point-like in the internal dimensions it is just a D3-brane and can be moved away from the bulk D7-brane \cite{Douglas:1995bn}. In this case, we end up with the $SU(2)^3$ quiver model,
a bulk brane with charge vector \eqref{ChVectDbZeroFlux} and a mobile D3-brane. If the D3-brane moves on top of the singularity, we obtain the $SU(3)^3$ quiver model and the bulk configuration we started with.

A summary of the different connected quiver gauge theories at the dP$_0$ singularity is shown in Figure~\ref{fig:theweb}.

\subsection{Local effective field theory description}

Having found a continuous connection between different quiver gauge theories,
a natural question to ask is if it is possible to describe this effect in terms of flat directions in field theory space.
In this section, we shall investigate only a possible local effective field theory description focusing just on states between the D3- and flavour D7-branes. A proper full understanding of these transitions would involve the inclusion
of bulk D7 states but the derivation of the global effective field theory (EFT) is beyond the scope of this paper.

Following our discussion of supersymmetric bulk-to-flavour brane transitions, we can immediately infer three important facts:
\bi
\item Transitions of type I involve brane creation/annihilation processes which are purely stringy effects. This makes an EFT interpretation more involved.

\item Transitions of type II  
proceed just through brane splitting/recombination processes,
corresponding to gauge enhancement/higgsing of the quiver gauge group. Thus we expect to be able to find a local EFT description
of these phenomena.

\item Combinations of transitions of type I and II which reduce or increase the rank of the gauge group can be interpreted in the local EFT
by appropriate fields switching their VEV on or off. However, given that some steps proceed via transitions of type I, the interpretation of these processes is less clear.
\ei

Let us first show how a transition of purely type II can be interpreted in the local EFT.
As an example, we shall consider the transition from $SU(3)\times SU(2)^2\times U(1)^2$ to $SU(2)^3$
which could be parameterised by the flat direction $J_{II} = A_1B_1C_1$ (in the absence of superpotential couplings
of the form $W\supset A_1B_1C_1$ which would break the F-flatness condition) with
\be
\label{flatII}
A_1=(3,1,1)_{-0}\,, \qquad B_1=(\bar{3},1,1)_{0+}\,,\qquad C_1=(1,1,1)_{+-}\,,
\ee
where the subscripts represent the $U(1)^2$ charges.
Notice that $A_1$ and $B_1$ are D3-D7 states whereas $C_1$ is a D7-D7 state.
The second ones are always present when the flux on the flavour brane is such that it makes the FI-term vanish (for shrunk dP$_0$).\footnote{Zero FI-term for the flavour branes means that the contribution on the flux from the bulk is suppressed.}

Let us now show how combinations of transitions of type I and II could also be described in terms of
flat directions in the local EFT. For example, the
$SU(3)^2\times SU(2)\times U(1)$ model can be connected to the $SU(2)^3$ one by the combination of one
transition of type I and another of type II. In this case the flat direction could be determined by the invariant $J_{I+II}=X_1 U_1 W_1$ with
(assuming again the correct superpotential couplings)
\be
\label{flat}
X_1=(3,\bar{3}, 1)_0\,, \qquad U_1= (\bar{3},1,1)_+\,,\qquad W_1=(1,3,1)_-\,,
\ee
where the subscripts represent again the $U(1)$ charges. The flat direction $\langle X_1\rangle=\diag(0,0,v_3)$,
$\langle U_1\rangle =\langle W_1\rangle =(0,0,v_3)$ gives precisely the spectrum of the $SU(2)^3$ quiver gauge theory.
Notice that $X_1$ is a D3-D3 state whereas $U_1$ and $W_1$ are D3-D7 states.

As a second example, let us consider the process of removing one D3-brane from the singularity (that can be seen as 
a combination of two transitions of type I with one transition of type II). This will
connect the $SU(3)^3$ model to the $SU(2)^3$ one. The flat directions can be easily identified as follows. The original spectrum can be written as
\be
\label{trinification}
X_i=(3,\bar{3},1)\,, \qquad Y_i=(\bar{3},1,3)\,, \qquad Z_i= (1,3,\bar{3})\,,
\ee
where $i=1,2,3$ is a family index. The flat directions can be taken to be $\langle X_1 \rangle=\langle Y_1\rangle =\langle Z_1\rangle = \diag(v_1, v_2, v_3)$. The existence of the gauge invariant $J_{I+I+II}=X_1Y_1Z_1$ guarantees this configuration to be D-flat. It is also F-flat given the general structure of the dP$_0$ superpotential $W=\epsilon_{ijk}X_iY_jZ_k$ since $\langle X_{2,3} \rangle= \langle Y_{2,3}\rangle = \langle Z_{2,3} \rangle =0$.  For $v_1=v_2=0$ the $SU(3)^3$ symmetry is broken to $SU(2)^3$ with the massless spectrum matching precisely that of the corresponding quiver.\footnote{The 81 states in $X_i,Y_i,Z_i$ split into the 36 states of the $SU(2)^3$ quiver and 45 massive states. 15 of these 45 states are eaten Goldstone modes from $X_1,Y_1,Z_1$ while the other 30 states, corresponding to the third row and column of the fields with zero VEVs, get a mass from the cubic superpotential.} Similarly, $v_2,v_3\neq 0$ gives the $U(1)^3$ quiver, and if also $v_1\neq 0$ the group is fully broken. These flat directions correspond to moving each set of three fractional branes out of the singularity to become one bulk D3-brane.

Regarding transitions of purely type I, like the one connecting the $SU(3)^3$ and the $SU(3)^2\times SU(2)\times U(1)$ models,
or the one relating the latter to the $SU(3)\times SU(2)^2\times U(1)^2$ quiver, we cannot interpret them as a simple Higgs mechanism as above since there are only bi-fundamentals in the spectrum and both groups have the same rank.
However one could still find a local EFT description of these processes by `going through a loop' in field space 
 considering different combinations of transitions of type I and II. For example, one could connect $SU(3)^3$ with $SU(3)^2\times SU(2)\times U(1)$ 
by going first from $SU(3)^3$ to $SU(2)^3$ along the flat direction (\ref{trinification}) and then from $SU(2)^3$ to $SU(3)^2\times SU(2)\times U(1)$
along (\ref{flat}).

We finally point out that there are some similarities with the case found in heterotic orbifolds~\cite{Ibanez:1987xa,Font:1988tp} in which one discrete Wilson line breaks an $SU(9)$ group to $SU(3)^3\times U(1)^2$ while a continuous Wilson line connects both models by first breaking $SU(9)$ to $SU(3)^3$ and then having a critical point in which the $SU(3)^3$ symmetry gets enhanced to $SU(3)^3\times U(1)^2$. However, an EFT flat direction only captures the breaking of $SU(9)$ to $SU(3)^3$ missing the enhanced symmetry point with the extra $U(1)^2$ symmetry. Moreover, starting from $SU(3)^3\times U(1)^2$ a flat direction gives rise to $SU(3)^3$, implying that this model interpolates between the two rank 8 enhanced symmetry points.

\section{Conclusions}
\label{sec:conclusions}

In this article we discovered new supersymmetric transitions which continuously connect at the classical level
four dimensional $\cN=1$ string vacua with different gauge group and chiral content. In particular,
we found that quiver gauge theories arising from D3/D7-branes at singularities which look completely
independent from the non-compact point of view, actually turn out to be all connected to each other
by considering the global embedding of these local models in compact CY manifolds.
This `web of quiver gauge theories' is parameterised by splitting/recombination modes of bulk D7-branes since
different gauge theories are connected via bulk-to-flavour brane transitions.

We described two types of basic transitions that can be used as building blocks for all the others: 
\bi
\item \emph{transitions of type I} which occur when a bulk D7-brane not passing through the singularity is deformed continuously 
until it touches the singularity and then combines with a fractional D3-brane to give a flavour D7-brane;

\item \emph{transitions of type II} which take place when a bulk D7-brane  
touches the singularity and then splits into a fractional D3-brane and two flavour D7-branes.
\ei
We stress that all these transitions are among different supersymmetric BPS configurations, and so they take place without any energy cost. 
Moreover, there is an upper bound on the number of possible transitions coming from the requirement of D7 tadpole cancellation 
which fixes the number of bulk D7-branes available for the transitions.

We illustrated these general claims in a particular example taken from our previous paper \cite{Cicoli:2013mpa}. There,
we constructed globally consistent chiral models 
with fractional D3-branes and flavour D7-branes within a compact CY manifold 
with an orientifold action that exchanges two dP$_0$ singularities. 
Within this context, we showed that by applying subsequently four transitions of type I, 
one can connect the $SU(3)^3$ trinification model to $SU(3)^2\times SU(2)$, then 
to the left-right symmetric gauge theory $SU(3)\times SU(2)^2$, to the Standard Model 
$SU(3)\times SU(2) \times U(1)$, and finally to $SU(3)\times U(1)^2$ where this is the last 
possible transition since D7-charge cancellation allows for only four bulk D7-branes.

We then took another route in our web of quiver gauge theories, and starting from the $SU(3)\times SU(2)^2$ model, 
we applied the inverse of a transition of type II which gives rise to the $SU(2)^3$ quiver with an additional 
bulk D7-brane supporting a non-Abelian flux. By a subsequent flux/brane transition a mobile D3-brane can be expelled from 
this bulk D7-brane into the whole CY manifold. From all this discussion, we learnt that the process of removing a D3-brane 
from the dP$_0$ singularity, for example going from $SU(3)^3$ to $SU(2)^3$, can also be realised by applying a sequence 
of transitions, two of type I and finally one of type II.
We also tried to interpret these continuous transitions in the EFT language in terms of VEVs of low-energy fields.

Our results open up several new avenues that would be interesting to explore in the future:
\bi
\item Besides understanding the kinematics of this web of quiver gauge theories, 
it would be very interesting to unveil the full dynamics which governs these transitions. 
Given that the flat directions of our web of quiver gauge theories are splitting/recombination modes of bulk D7-branes, 
we expect this dynamics to be determined by the open string potential within a moduli stabilisation setting similar to the one discussed in \cite{Cicoli:2013mpa}. In particular, these flat directions can possibly be lifted by supersymmetry breaking effects. In this case, we would be able to understand the dynamics of these transitions which is a crucial issue for addressing phenomenological questions such as why there are three families or why we observe the Standard Model gauge group.

\item Related to the previous issue, is the attempt to find a complete effective field theory description of these transitions 
in terms of VEVs in field space. This might not be entirely possible since some effects seem to be purely stringy, but a key question 
to answer to have a clearer picture is what is the structure of D3-D7 couplings.

\item An important further extension of our work is the study of these transitions in other geometric backgrounds 
such as higher del Pezzo singularities. Moreover, it would be interesting to analyse the interplay between bulk-to-flavour brane transitions 
in a compact embedding and transitions among different del Pezzo surfaces~\cite{Feng:2002fv} and Seiberg/toric duality~\cite{Feng:2001bn}.

\item On the cosmology side, having this richness of flat directions immediately suggests 
potential applications for inflation after moduli stabilisation is properly implemented. The fact that a brane/anti-brane annihilation interpretation enters in this scenario already within a supersymmetric setting may eventually be related to previous attempts of brane/anti-brane inflation including post-inflationary effects like the generation of topological defects such as cosmic strings at the end of inflation~\cite{Quevedo:2002xw,McAllister:2007bg,Cicoli:2011zz,Burgess:2011fa}. However, this web of quiver gauge theories may also give rise to new inflationary scenarios. The fact that the full structure of the flat directions is not properly captured by an EFT description may open the possibility of a truly stringy scenario for inflation.

\item A final interesting question to ask is to how these supersymmetric transitions can be interpreted when up-lifting these constructions to F-theory \cite{Collinucci:2008zs,Collinucci:2009uh,Blumenhagen:2009up}. 
Notice that so far there has been little work regarding F-theory constructions with singularities on the base since most of the work has been concentrated on singularities on the elliptic fibration (see however Section~4.3 of \cite{Aldazabal:2000sa} and \cite{Balasubramanian:2012wd}).
\ei

\subsection*{Acknowledgement}

We would like to thank Andres Collinucci, Shanta de Alwis, I\~naki Garc\'ia-Etxebarria, Noppadol Mekareeya and Angel Uranga
for useful discussions. The work of CM and SK was supported by the DFG under TR33 ``The Dark Universe''. SK was also supported by the European Union 7th network program Unification in the LHC era (PITN-GA-2009-237920). SK and CM would like to thank ICTP for hospitality.

\appendix

\section{Fluxes from non-complete intersections}
\label{AppFlux}

In this appendix, we construct an explicit two-form flux on the world-volume of the bulk D7-branes $\cD$ which is not the pullback of a two-form of the CY $X$. This means that the Poincar\'e dual two-cycle in $\cD$ is not described by one equation intersected with the D-brane equation and the CY equation. Moreover, we want that this curve is algebraic, such that the corresponding two-form flux is of type $(1,1)$. We follow the procedure outlined in \cite{Bianchi:2011qh}. To describe such curves we focus, for convenience, on one particular subset of the complex structure moduli space of $X$: take the equation \eqref{EqCY3} defining the CY $X$, and make the restriction $P^+_9=\tilde{P}_3\cdot \tilde{P}_6$. Then we can rewrite the hypersurface equation of the CY as\footnote{We have set $P^-_i\equiv 0$ in order to obtain an orientifold invariant CY.}
\begin{eqnarray}
\label{EqCY3Modif}
 eq_X &\equiv& z_5 \cdot Q_5 + z_6 \cdot Q_6 +\tilde{P}_3 \cdot Q_3 =0\:,
\end{eqnarray}
where $Q_i$ are polynomial of the coordinates different from $z_5$ and $z_6$. Consider now the following curve in the ambient four-fold $Y_4$
\begin{equation}
    \cC\::\qquad z_5=0 \quad \cap \quad z_6=0 \quad \cap \quad \tilde{P}_3=0  \:.
\end{equation}
These three equations automatically solve the CY equation \eqref{EqCY3Modif}. Hence, the curve $\cC$ is inside $X$. Next, take the equation \eqref{DbDiv} defining the D-brane wrapping $\cD_b$:
\begin{equation}\label{DbDivApp}
 z_4z_5 + \alpha z_6z_7 + P_3^{D7}\,z_4z_7z_8=0 \:.
\end{equation}
If $P_3^{D7}\equiv \tilde{P}_3$, then the curve $\cC$ lies also on the D-brane world-volume; its Poincar\'e dual two form $\omega^\cC$ is holomorphic and is of the type we are looking for. If we deform the D-brane equation by modifying $P_3^{D7}$,  the curve will be modified too and it will not be represented by the algebraic curve $\cC$ anymore. Hence, it can develop  $(0,2)$ components. Such a
`deformed' flux would break supersymmetry. The flux $\omega^\cC$ fixes, therefore, the deformations which would cause $(0,2)$ components.

Let us see what the homology class of the curve $\cC$ in the ambient four-fold is:
\begin{equation}\label{cCclassCY}
 [\cC] = D_5\cdot D_6 \cdot 3D_1 = (D_5+D_4)\cdot (D_6+D_7) \cdot \tfrac13 [X] = \tfrac13\cD_b \cdot \cD_b\cdot [X] =D_1 \cdot \cD_b\cdot [X]\:.
\end{equation}
Hence, the two-cycle $\cC-D_1\cdot \cD_b$ is trivial on the CY three-fold,\footnote{In the example we are considering, the two cycles homologous in the ambient four-fold are homologous on $X$ as well.} but the two-form $\omega^\cC-D_1$ is non-trivial on the divisor $\cD_b$.

Now we go to the limit $\alpha\rightarrow 0$, when the D-brane splits into a brane wrapping the dP$_0$ and one flavour brane with equation
\begin{equation}\label{FlavDbDivApp}
 z_5 +  P_3^{D7}\,z_7z_8=0 \:.
\end{equation}
If $P_3^{D7}\equiv \tilde{P}_3$, the curve $\cC$ is also inside the flavour D7-brane. The same consideration made above are valid for the Poincar\'e dual two-form flux $\omega^\cC_{\rm flav}$.

To compute the D3-charge of the flux, we need to compute its square $\int_{\cD_b} \omega^\cC\wedge \omega^\cC$. Since its Poincar\'e dual two-cycle is not a complete intersection with the D-brane equations we have to use the following relation:\footnote{See \cite{Braun:2011zm,Louis:2012nb} for applications of the same trick in an analogous context and for more details.}
\begin{equation}
 \int_{\cD_b} \omega^\cC\wedge \omega^\cC = \int_\cC \omega^\cC=\int_\cC c_1(N|_{\cC\subset \cD_b})\:.
\end{equation}
The last equality uses the fact that the Poincar\'e dual of a curve in a surface is equal to the first Chern class of the normal bundle of that curve in the surface. From the following exact sequence
\begin{equation}
 0\rightarrow N_{\cC\subset \cD_b} \rightarrow N_{\cC\subset Y_4}\rightarrow N_{\cD_b\subset Y_4}  \rightarrow 0 \:,
\end{equation}
we find
\begin{eqnarray}
 c_1(N|_{\cC\subset \cD_b}) &=& c_1(N|_{\cC\subset Y_4})-c_1(N|_{\cD_b\subset Y_4}) \nonumber\\
 				&=&  \left( D_5+D_6+3D_1\right) - \left(2 D_5 + D_6 + D_4 + D_5+D_4  \right)\nonumber\\
 				&=& 3 D_1 - 2 D_5 - 2 D_4 \:.
\end{eqnarray}
Hence,
\begin{eqnarray}
 \int_{\cD_b} \omega^\cC\wedge \omega^\cC &=& \int_\cC c_1(N|_{\cC\subset \cD_b}) \nonumber\\
 			&=& \int_{Y_4} D_5 \cdot D_6 \cdot 3D_1 \cdot (3 D_1 - 2 D_5 - 2 D_4) = -9 \:.
\end{eqnarray}
By using this result and the relation \eqref{cCclassCY}, which implies $\omega^\cC\cdot D_1|_{\cD_b}=D_1\cdot D_1|_{\cD_b}$, we can compute the square of $\cF_b=\omega^\cC-D_1$:
\begin{eqnarray}
  \int_{\cD_b}\cF_b^2 &=& \int_{\cD_b} \left(\omega^\cC-D_1\right)^2 = \int_{\cD_b} \left(\omega^\cC\right)^2 - \int_{\cD_b} D_1^2 = -9-3=-12
\end{eqnarray}

We obtain the same results for the flux $\omega^\cC_{\rm flav}$, i.e.\ $\int_{\cD_b-\cD_{q_1}}\omega^\cC_{\rm flav}\wedge \omega^\cC_{\rm flav}=-9$.
Moreover, from the SR-ideal of $Y_4$, one can see that $\cC\cdot \cD_{q_1}=0$.

\begin{footnotesize}
\bibliographystyle{utphys}
\bibliography{CKMQVnew}
\end{footnotesize}
\end{document}